# Innovation in Large-scale agile - Benefits and Challenges of Hackathons when Hacking from Home

Rasmus Ulfsnes[1][0000-0002-4966-8242], Viktoria Stray[1,2][0000-0002-6032-2074], Nils Brede Moe[1][0000-0003-2669-0778] and Darja Šmite[3, 1][0000-1111-2222-3333]

[1] SINTEF, Trondheim, Norway
`{rasmus.ulfsnes, nils.b.moe}@sintef.no`
[2] Department of Informatics, University of Oslo, Oslo, Norway
`stray@ifi.uio.no`
[3] Blekinge Institute of Technology, Karlskrona, Sweden
`darja.smite@bth.se`

**Abstract.** Hackathons are events in which diverse teams work together to explore, and develop solutions, software or even ideas. Hackathons have been recognized not only as public events for hacking, but also as a corporate mechanism for innovation. Hackathons are a way for established companies to achieve increased employee wellbeing as well as being a curator for innovation and developing new products. Sudden transition to the work-from-home mode caused by the COVID-19 pandemic first put many corporate events requiring collocation, such as hackathons, temporarily on hold and then motivated companies to find ways to hold these events virtually. In this paper, we report our findings from investigating hackathons in the context of a large agile company by first exploring the general benefits and challenges of hackathons and then trying to understand how they were affected by the virtual setup. We conducted nine interviews, surveyed 23 employees and analyzed a hackathon demo. We found that hackathons provide both individual and organizational benefits of innovation, personal interests, and acquiring new skills and competences. Several challenges such as added stress due to stopping the regular work, employees fearing not having enough contribution to deliver and potential mismatch between individual and organizational goals were also found. With respect to the virtual setup, we found that virtual hackathons are not diminishing the innovation benefits, however, some negative effect surfaced on the social and networking side.

## 1    Introduction

Innovation in large companies is important but is often harder to implement than in start-ups where it is a natural and necessary part of the regular work. Large companies have a high number of employees, many of them creative, but teams and team members are often bounded to work only on the company strategy [1]. Creating an environment that fosters innovation and creativity requires employees to be motivated, and management needs to provide time and space for innovation to occur [1]. Therefore, to address the challenge of bringing new products to the market, many large software companies tend to approach innovation in a systematic way [1]. One example of such approach is implementing an innovation program for internal startups [2]. Google and Atlassian had their 20% time program [3], where developers are given 20% of their time to work on a project, or initiative of their own choosing. Another approach is to allocate days when all software developers in a company work on delivering a software product improvement of their choice. This can be facilitated in a hackathon, a short and time bound event, where participants work together to explore, and develop solutions, software or even ideas. Large-Scale Agile organizations have utilized hackathons increasingly as a mechanism for innovation in the last 20 years [3]–[5].

Due to Covid19, company activities such as hackathons had been first postponed and eventually organized through virtual platforms such as MS Teams, Zoom, and Slack. Further, as many companies have announced their *Work from anywhere* (WFA), strategy (Twitter, Spotify, Facebook, Salesforce), it is likely that structured innovative processes like hackathons will be organized virtually in the future too.

This unique situation has provided an excellent opportunity to understand not only how the shift from physical to a virtual hackathon affects the benefits and challenges of hackathons, but also how large-scale agile organizations can embrace a WFA future. Even though there has been an increasing amount of literature on hackathons [7], and some literature on virtual hackathons [6] the literature regarding virtual hackathons for software companies, and especially the shift from physical to virtual, is scarce.

Motivated by the importance of innovation in large-scale agile and how large-scale agile organizations can embrace the WFA future, our research questions are:

**RQ1: What are the benefits and challenges of hackathons?**

**RQ2: How are benefits and challenges affected by moving to a virtual hackathon?**

Our chosen way to investigate this question was to perform a case study in a multinational software company utilizing large-scale agile methodologies. We observed the company employees during their virtual hackathon and inquired about the recognized benefits and challenges of hackathons for the company and their employees, and how the virtual hackathon compared to the previous physical hackathons.

## 2    Background

Hackathons have been around for over 20 years, first appearing with OpenBSD and SUN microsystems in 1999 [8]. The word originates from the combination of the words "hack" and "marathon" [8]. There are multiple alternative names for hackathons, such as hackfest, jam, codefest, bug bash [5], as well as other more obscure names such as Delivery Day [4] and FedEx Day [3]. These events all contain similar elements, such as a limited and defined amount of time and the goal to create some kind of minimum viable product (or at least something to show at a demo). Hackathons can be open events organized by universities, cities, municipalities, or internal corporate events.

Hackathons vary in how they are organized and executed, how ideas are structured, where it takes place (physical or virtual), and whether or not it is a competition [6]. These characteristics of hackathons provide different benefits and challenges. Falk Olesen and Halskov [7] emphasize that hackathon organizers need to tune its characteristics in order to achieve the wanted benefit for the target group.

Hackathons are not only a benefit for the organization, but also for the individual participants, providing opportunity for individual development, goals and learning skills. There are also different categories of benefits that are identified in hackathons. Falk Olesen and Halskov [7] provides three categories, *structure learning, structuring process and enabling participation.*

While hackathons provide benefits such as improving and fostering internal innovation, expanding competencies, and networking [3], [4], [6], [9]–[12], several challenges have been reported. Examples of challenges are prototypes that do not get sufficient follow-up work [7], [9], [12], stress by putting regular work aside [4] and issues associated with *hacker culture* and *toxic masculinity* deferring especially women [10].

Contradictions and tradeoffs are something apparent in the hackathon phenomenon, regardless of its formats. Diversification is both a benefit for learning and networking and a potential hinder for effective work during the hackathon [9]. Another contradiction is that the individual participants want to learn one specific technology or skill, disregarding whether or not that is useful for the organization [13].

## 3   Research methodology and approach

Our exploratory case study was conducted in a multinational agile software company identified here as "Ares" due to confidentiality. The company was founded in the early 2000s and develops content distribution software for mobile devices. Ares's developers are located in two locations and are utilizing agile methodologies. Ares organizes a voluntary hackathon every year. Ideas are submitted, and put up for voting based on managerial input. The employees then vote for three ideas they would like to work on. The teams are constructed based on the votes and managerial input to ensure good team sizes and diversity. The hackathon lasts two days, teams are self-organized, and at the end, all teams present a demo of their work. There are no awards or jury involved.

In this case study, we first analyzed a demo from a recent hackathon in the company. Based on the demo and previous research [3], [4], we developed an interview guide containing different areas for questions: *Background, Motivation, Idea generation, Co-operation, Hackathon organization, Expectations, Benefits, Challenges, Virtual.*

Eight semi-structured interviews with people from four different teams were conducted. They represented different skills and roles within the company, including developers, lead architects, senior engineers, and an advertisements operations manager. The interviews were conducted in Norwegian or English, recorded, transcribed and qualitatively coded using descriptive coding, followed by a holistic overview and a thematic analysis [13]. Finally, to validate the qualitative findings and elicit additional data, we conducted a survey with 23 respondents.

## 4 Results

In this chapter, we present our results by starting with the characteristics of hackathon, which provide a context for understanding our findings.

**How often? – Once a year.** To our surprise, even though the participants were very positive towards hackathons, they did not want to conduct them too often. One participant explained, "*It's such a nice activity it's something like your birthday. If it would be every month, it would become a bit more boring. It would be just another Hackathon.*"

**How long? – Time boxed**. This is an inherent part of the hackathon, which often means that participants do not have time to finish the project. Opinions about time boxing differ. As one states, "*You should stop working when you are finished with the hackathon. Either you prove that something works, or you hopefully had fun.*" This is in contrast to those who said that continuation of the project was important to them.

**Which ideas? – Whichever.** The freedom to pitch and develop any idea gives a motivational boost and incentive to participate. As one explains, "*... with a more open approach, you can create ideas which might sound cool, but you can't really integrate it into your company. However, on the other side, those ideas might be something new which your company could work on.*" Also, the opportunity to explore personal interests was important, one example being the idea of developing a guitar tuner popular among guitar players.

**With whom? – Whomever.** The informants noted that the learning and team building across disciplines and skills was a vital aspect of participating in the hackathon. In other words, indicating that a diverse team, with different skill-levels was good. This however was contrasted by the importance for the participants to succeed. One of the participants expressed, "*If I constructed the teams, I probably would have chosen different people*"

**What's at the end? – Demo, no competition.** One informant explained, "The *demo is everything about the hackathon, the hackathon is not successful if you do not have something to show at the demo.*" That being said, the focus on the demo can also make some people reluctant to join, as one stated: "*I haven't participated in hackathons earlier. I worried that I would not contribute enough because you need to deliver a product and present it at the end.*"

### 4.1 Benefits of Hackathons in Large-Scale Agile

Based on the interviews we will now present the general benefits of hackathons categorized into individual and organizational. We report the benefits of most importance, based on the rating provided by the survey respondents (see Table 1).

**Table 1** Rated benefits of hackathons

| # | Individual - Benefits | Organizational - Benefits |
|---|---|---|
| 1 | Test new ideas fast | Teambuilding |
| 2 | Break from ordinary work | Ability to take advantage of emergent market opportunities |
| 3 | Fun from being with colleagues | Build a sense of belonging to the company |
| 4 | Build new stuff | Employees with broader skills and expertise |
| 5 | Build a network with colleagues | Increased marketability and talent attractiveness |
| 6 | Expand skills and expertise | Knowing who knows what (build employee network) |

Our findings show that innovation is a key benefit both for the individuals and for the organization, recognized as the ability to test new ideas fast and to take advantage of emerging market opportunities. One informant said, *"It was a product idea I had wanted to develop for a while and finally there was a hackathon so I could check if it worked."* Other interviewees also mentioned that given the right business case, hackathon products could be launched. One explained: *"Two years ago, we made a feature that was put directly into production, actually four weeks later, it was done properly. Hackathon proved it was possible and it fit well into the existing product line."*

Having a break from ordinary work was highly valued by individuals. One interviewee explained: *"It's kind of relaxing for the people, for the company. Well for the team itself because you are not doing exactly what you are paid for."*

Teambuilding was rated as the top organizational benefit and having fun from being with colleagues was rated third as the individual benefit. One informant explained, *"In hackathons, what normally happens is that the main outcome, at least in our experience, tends to be the social aspect of it, the teambuilding."* Another stated, *"you will remember other important parts of the hackathon than just the work or who was the winner, for example, the fun you had with your colleagues and the knowledge you gained."*

Increased marketability and talent attractiveness are achieved due to the individual benefits, but also because the hackathon in itself is a company practice that gives an edge during the recruitment process. The individual expands their skills and expertise by working on topics and technology where the participant did not have much prior knowledge and thus utilizing the hackathon for learning, providing the organization with higher and broader skilled employees. Building network among colleagues and having employees that know who knows what increase the inter-team communication and give employees new insights into and respect for one another's role and work.

### 4.2    Challenges of Hackathons in Large-Scale Agile

Even though there are clear benefits of hackathons, employees also mentioned some challenges, we mention the challenges corroborated by the survey. The complete list of challenges is shown in Table 2.

The top individual challenge was related to hackathons adds stress. Even though hackathons are mostly a welcomed break, the timing for running a hackathon is not always good. As someone explained, *"you have pressure during your workday, and you need to get something finished and you look at hackathon as something in the way that is just disturbing you from important work."* This challenge is echoed two challenges on the organizational side – stopped production for a time and increased employee stress.

Mismatch in skills and desires due to the team diversification was another top challenge. The optimal team structure was not always obvious. One participant described that the team members did not always have the same desires and motivation for participating in the hackathon. One interviewee explained, *"It was challenging that we had different goals we wanted to achieve. I wanted my product to be launched while the others wanted to expand their skills on machine learning."*

Table 2. Challenges of hackathons

| Individual - Challenges | Organization - Challenges |
|---|---|
| Overhead of hackathon adds additional stress | Stops production for a time |
| Mismatch in skills leads to slower development | Employees do not partake in hackathon benefits |
| Being unable to complete projects is somewhat frustrating | Mismatch in skills leads to less time spent on emerging products |
| Lack of confidence is a barrier for joining | No-continuation plan leads to discontinued projects |
| Mismatch in individual goals for learning leads to disagreement in the teams | Employees are acquiring non-relevant skills |
| Bad communication and coordination due to mismatch in individual goals | Increased employee stress |

Being unable to complete the hackathon projects is somewhat frustrating. Even though the employees are well aware that the hackathon is timeboxed, they still have a desire to finish their products. A participant explained, "*I am not allowed to work on it, I have other work tasks. That's a downside with hackathons.*" This results in developers feeling dissatisfied as their goals remain unfulfilled, as someone explained, "*You know, the worst feeling you can have is when you write code and after some time it gets forgotten and no one ever uses it.*"

**Virtual Hackathons – What are the Changes?**

The ability to take advantage of emerging market opportunities achieved through hackathons are not impacted by a virtual hackathon. The idea generation and idea voting are similar to the previous hackathons, the only difference is that the teams work and collaborate virtually. As one says, *"But from a technical point of view, to just develop something, you can always use Slack or whatever, or just Google meet and call."* In addition to this, our findings show that the number of hours spent on virtual hackathon is less than when it is held physically. "*During the first hackathon we stayed in the office up until midnight.*" In contrast, in the virtual hackathon, the employees started by dividing the work, then worked focused individually, continuously coordinating through Slack and using virtual meeting rooms for discussions and planning: *"We did the first meeting when we split the work. So, after a few hours we met again. We have this approach, let's just give it a try. Then a few hours later we had another meeting, and we tracked our progress quite often."*

A break form ordinary work through hackathons is especially welcome after stressful periods such as migrations, or as one says in a pandemic," Especially *in these pandemic times. You know we are working from home. So, we want to meet people in real life and hackathons are one way to just, you know, meeting and collaborating"*.

We see a reduction in the benefit of having fun with colleagues: "*It was not as fun as it used to, hackathons are usually more fun than over video during Covid, in my own home.*" Most employees mentioned that they would prefer to have the hackathon in the same physical location in order to socialize and have a good time. Lastly, disagreements due to mismatch in individual goals seems to be harder to solve through virtual meetings.

## 5 Discussion

Noting the lacking research on internal corporate hackathons [9], we will now discuss our research questions. Our first research question was "What are the benefits and challenges of hackathons?" Consonant with [3] and [4], we found that providing dedicated days for developers may lead to innovation and new product development, and that there are benefits both on the individual and company levels as seen in **Table 1**. Developer involvement, productivity, competence and well-being are thus important factors for large-scale agile companies that foster innovation. We also found that learning and networking are important benefits, as corroborated by [7], [10]. In addition, just providing time to take a break from regular work and having fun is important for the well-being of the employees. In addition to the different levels of benefits, we also see a clear link between the characteristics of hackathons and the benefits that emerge from them. We also find tradeoffs between some of the benefits. For example, building new stuff vs teambuilding as our findings show that tailoring the teams for diversification might take away the team's ability to perform during the hackathon, corroborating Falk Olesen and Halskov [7] where the skills wanted by the individual did not match that of the organization. Our findings showed that some employees were frustrated when their product was not continued, this confirms a challenge with hackathon continuation as also reported in previous research [5], [12]. We also found additional challenges such taking a time away from everyday work, whether it was good timing or not, corroborated by [4]. Personal desires and goals could also pose a challenge where teams could not agree on common goals.

Our second research question was "How are benefits and challenges affected by moving to a virtual hackathon?" Our results suggest that the shift from physical to virtual hackathons in large-scale agile does not seem to change the innovation benefits or worsen the challenges that much. We found that the number of hours spent on the hackathon decreased compared to physical hackathons, and the work was more divided and focused between the participants. The employees were producing good quality demo prototypes and were happy with the outcome. The findings suggest that having fun with colleagues has decreased in virtual hackathons, since physical hackathons provide more encouragement and room for socializing, which was not provided in the virtual hackathon.

## 6 Conclusion and future work

Our results suggest that virtual hackathons in large-scale agile organizations benefits to the individuals' opportunity to test new ideas, take a break from regular work and acquire new skills, and provide the organization with new products, refreshed employees and a more competent workforce. However, practitioners should be aware that challenges can include additional stress, and dissonance in hackathon teams due to differences in experience and personal goals.

Future research should focus on collecting more data about virtual hackathons in other large-scale agile organizations to further develop the insights into how virtual hackathons can be organized and improved. When doing this, one could consider using the SPACE framework for developer productivity suggested by Forsgren et al. [14]. The framework recognizes multiple productivity dimensions for large-scale agile organizations, both on an individual, team and organizational level that are similar to our findings on hackathons.